\documentclass[final,1p,times]{elsarticle} % do not change
\usepackage{graphicx} % do not change
\usepackage{amssymb} % do not change
\usepackage{amsthm} % do not change
\journal{Nuclear Physics A} % do not change
\begin{document} % do not change

\begin{frontmatter} % do not change

%% QM09Author: please enter your  
%% Title, author and address info here; please do not use footnotes

% Your Title - please modify
%\title{EPS09 --- Nuclear PDFs and Their Uncertainties at NLO}
\title{EPS09 --- Global NLO Analysis of Nuclear PDFs and Their Uncertainties}

% Principle author, and co-authors - please modify
\author{Kari J. Eskola$^{a}$}
\author{Hannu Paukkunen$^{a}$}
\author{Carlos A. Salgado$^{b}$}

% Address - please modify
% note that if you have authors from several institutions, we recommend
% labelling these [a], [b], [c] etc.

\address[a]{University of Jyv\"askyl\"a,
Department of Physics, and Helsinki Institute of Physics, Finland}

\address[b]{Departamento de F\'\i sica de Part\'\i culas and IGFAE, Universidade de Santiago de Compostela, Spain}

\begin{abstract} % do not change

In this talk, we present our recent next-to-leading order (NLO) nuclear parton distribution functions (nPDFs), which we call EPS09. As an extension to earlier NLO analyses, we supplement the deep inelastic scattering and Drell-Yan dilepton data by inclusive midrapidity pion measurements from RHIC in order to reduce the otherwize large freedom in the nuclear gluon densities. Our Hessian-type error analysis leading to a collection of nPDF error sets, is the first of its kind among the nPDF analyses.

\end{abstract} % do not change

\end{frontmatter} % do not change

%% QM09: we keep linenumbers at least for initial version
%\linenumbers % do not change
\section{Introduction}
\vspace{-0.2cm}
The global analyses of the free nucleon parton distribution functions (PDFs) are grounded on the asymptotic freedom of QCD, factorization and parton evolution. These features allow to express the hard-process cross-sections formally as
$$
\sigma_{AB\rightarrow h+X} = \sum_{i,j} f_{i}^A(Q^2) \otimes \hat{\sigma}_{ij\rightarrow h+X} \otimes f_{j}^B(Q^2),
$$
where $f_{i}$s are the scale-dependent PDFs, and $\hat{\sigma}_{ij\rightarrow h+X}$ denote the perturbatively computable partonic pieces. The factorization theorem has turned out to work extremely well with increasingly many different types of data included in the latest free proton PDF analyses. For bound nucleons factorization is not as well-established, but it has anyway proven to do a very good job \cite{Eskola:1998iy,Eskola:1998df,Eskola:2007my,Eskola:2008ca,Hirai:2007sx,deFlorian:2003qf} in describing the measured nuclear modifications $\sigma_{\rm bound}/\sigma_{\rm free}$ in deep inelastic scattering (DIS) and Drell-Yan (DY) dilepton production involving nuclear targets. This talk summarizes our new NLO analysis \cite{Eskola:2009uj}.
\vspace{-0.2cm}
\section{Analysis Method and Framework}
\vspace{-0.2cm}
We define the nuclear PDFs through nuclear modification factors $R_i^A(x,Q_0^2)$ as follows:
$$
f_{i}^A(x,Q^2) \equiv R_{i}^A(x,Q^2) f_{i}^{\rm CTEQ6.1M}(x,Q^2),
$$
where $f_{i}^{\rm CTEQ6.1M}(x,Q^2)$ refers to a CTEQ set of the free proton PDFs \cite{Stump:2003yu} in the zero-mass variable flavour number scheme. The $x$ and $A$ dependences of $R_i^A(x,Q_0^2)$s satisfying the sum rules, are parametrized considering three modifications: $R_V^A(x,Q_0^2)$ for both valence quarks, $R_S^A(x,Q_0^2)$ for all sea quarks, and $R_G^A(x,Q_0^2)$ for gluons. The nuclear PDFs at $Q^2 > Q_0^2$ are obtained as solutions to the DGLAP equations using our own NLO DGLAP solver based on a semi-analytical method described e.g. in \cite{Santorelli:1998yt,Paukkunen:2009ks}.

All cross sections are computed using the factorization theorem and the initial parametrization is adjusted to find the minimum of 
\begin{eqnarray}
\chi^2(\{a\})    \equiv  \sum _N w_N \, \chi^2_N(\{a\}), \,\, {\rm where} \quad
\chi^2_N(\{a\})  \equiv  \left( \frac{1-f_N}{\sigma_N^{\rm norm}} \right)^2 + \sum_{i \in N}
\left[\frac{ f_N D_i - T_i(\{a\})}{\sigma_i}\right]^2. \nonumber
\end{eqnarray}
For each data set $N$, the $D_i$ denotes the experimental data value with point-to-point uncertainty $\sigma_i$, and $T_i$ is the theory prediction computed using parameters $\{a\}$. The pion data suffers from an overall $\sim 10\%$ normalization uncertainty $\sigma_N^{\rm norm}$, and the normalization factor $f_N \in [1-\sigma_N^{\rm norm},1+\sigma_N^{\rm norm}]$ is determined by minimizing $\chi^2_N$. The weight factors $w_N$ amplify the importance of those data sets whose content is physically relevant, but whose contribution to $\chi^2$ would otherwize be too small to be noticed by the automated minimization routine we use.

Besides finding the central set of parameters $S^0 \equiv \{a^0\}$ that optimally fits the data, propagating the experimental uncertainties to the PDFs has become an inseparable part of the modern PDF fits. The Hessian method \cite{Pumplin:2001ct}, which we use, trusts on a quadratic approximation 
$$
\chi^2 \approx \chi^2_0 + \sum_{ij} \frac{1}{2} \frac{\partial^2 \chi^2}{\partial a_i \partial a_j} (a_i-a_i^0)(a_j-a_j^0) \equiv \chi^2_0 +  \sum_{ij} H_{ij}(a_i-a_i^0)(a_j-a_j^0).
$$
The eigenvectors $z_k$ of the Hessian matrix $H$ serve as an uncorrelated basis for building the PDF error sets $S_k^\pm$. These are obtained by displacing the fit parameters to the positive/negative direction along $z_k$ such that $\chi^2$ grows by a certain amount $\Delta \chi^2$ from the minimum $\Delta \chi^2_0$. Using these sets, the upper and lower uncertainty of a quantity $X$ can be written e.g. as
\begin{eqnarray}
(\Delta X^\pm)^2 & \approx & \sum_k \left[ \max / \min \left\{ X(S^+_k)-X(S^0), X(S^-_k)-X(S^0),0 \right\} \right]^2 \label{eq:error_best}
\end{eqnarray}
where $X(S^\pm_k)$ denotes the value of the quantity $X$ computed using the set $S_k^\pm$. Requiring each data set to remain close to its 90\%-confidence range, we obtain $\Delta \chi^2=50$.
\vspace{-0.2cm}
\section{Results and Conclusions}
\vspace{-0.2cm}
\begin{figure}[!htb]
\center
\includegraphics[scale=0.4]{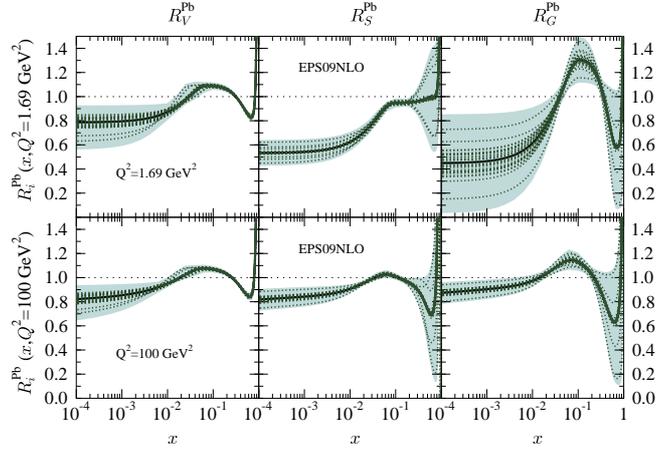}
\caption{The obtained modifications $R_G^{\rm Pb}$ at $Q^2_0=1.69 \, {\rm GeV}^2$ and at $Q^2=100 \, {\rm GeV}^2$. The black lines indicate the best-fit, whereas the dotted green curves denote the individual error sets which combine to the shaded bands like in Eq.~(\ref{eq:error_best}).}
\label{Fig:PbPDFs}
\end{figure}
Now, we briefly summarize the main results from the present analysis, starting with Fig.~\ref{Fig:PbPDFs} where we show the obtained modifications for Lead --- the nucleus relevant for the LHC --- at two scales. Interestingly, even the rather large uncertainty at small-$x$ gluons becomes notably smaller in the scale evolution. This is a prediction that might be testable in the future colliders. 

\begin{figure}[!htb]
\center
%\vspace{-0.2cm}
\includegraphics[scale=0.27]{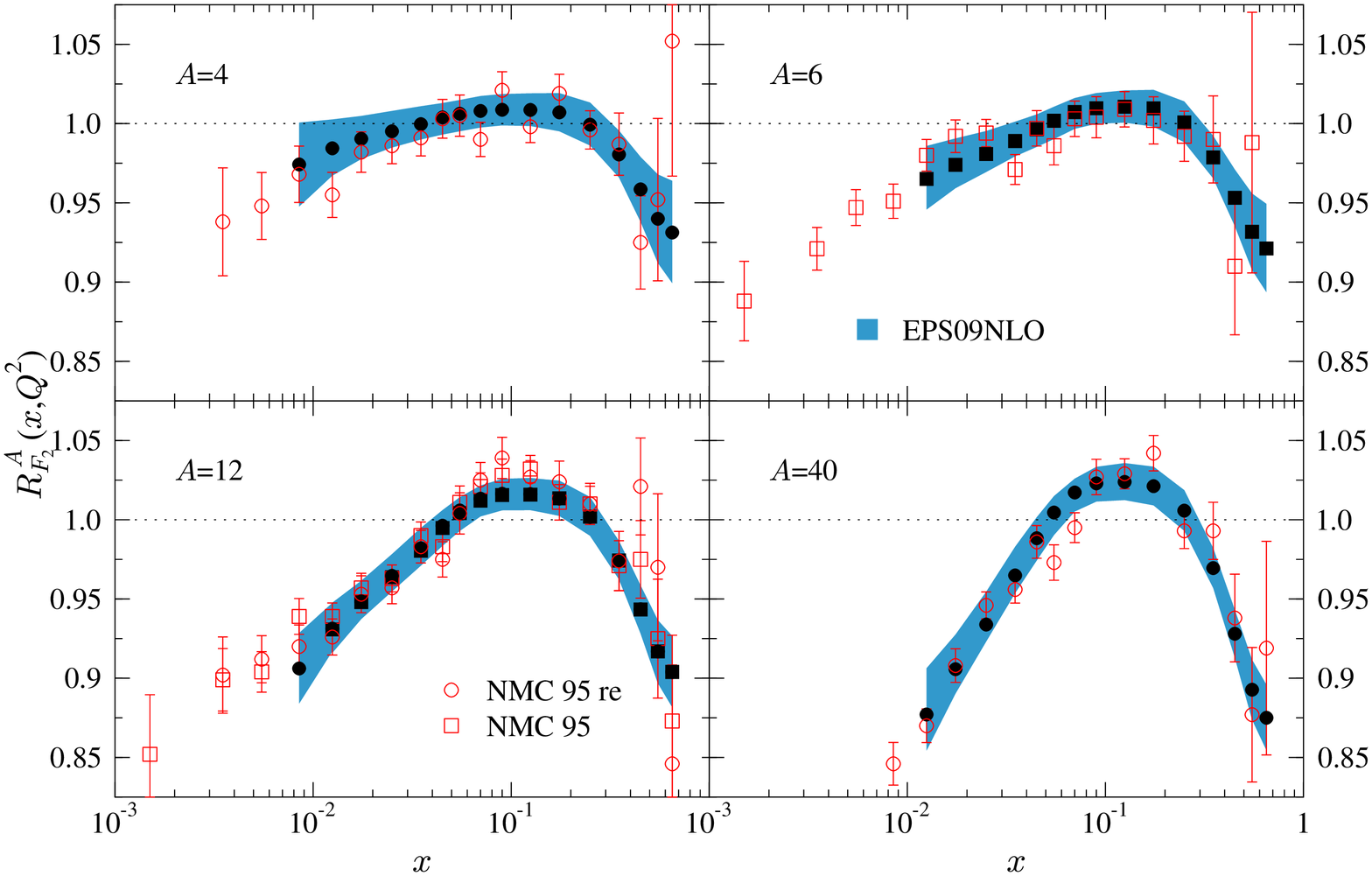}
\hspace{-0.5cm}
\includegraphics[scale=0.28]{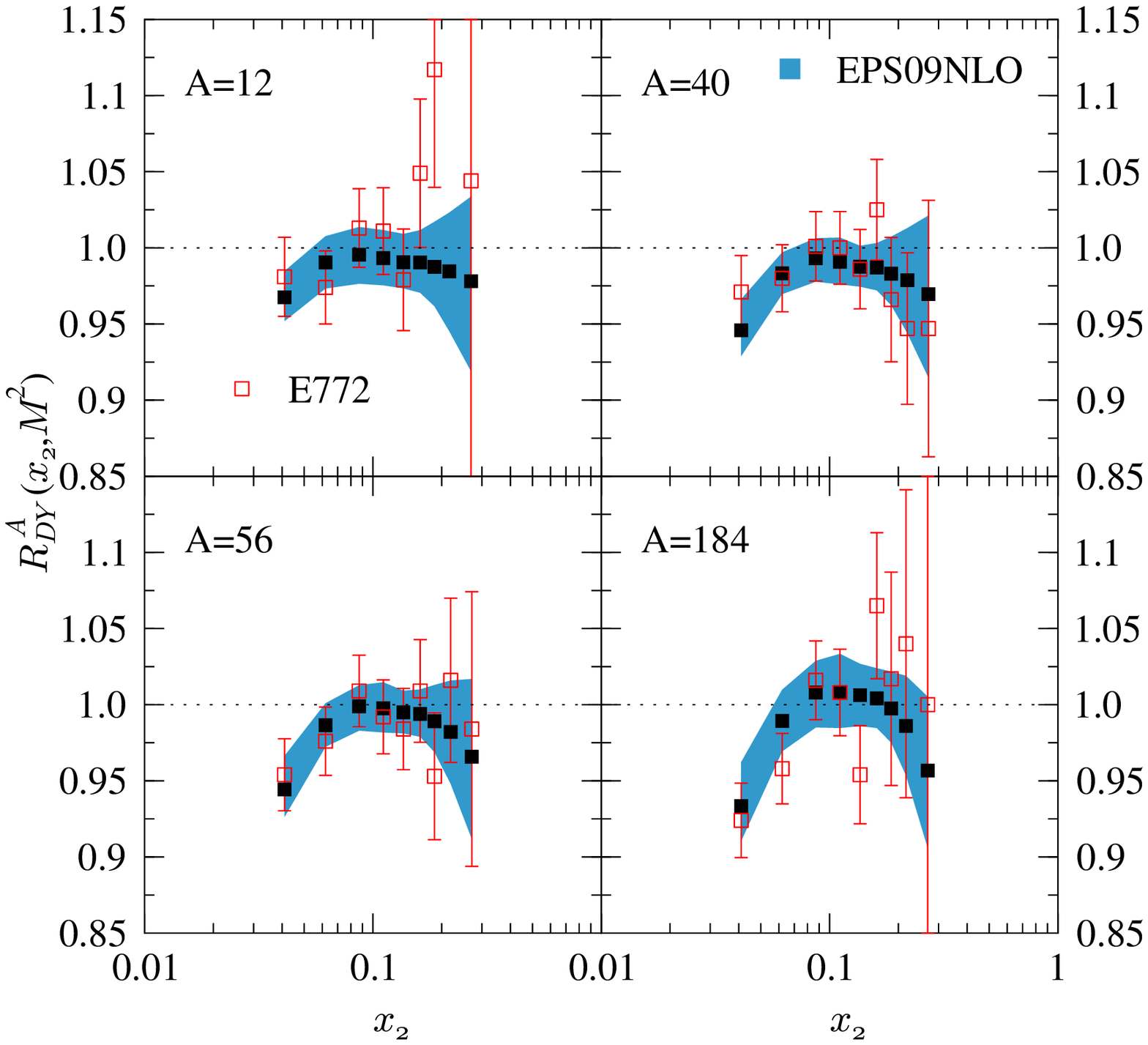}
\caption{The calculated $R_{F_2}^A$ and $R_{\rm DY}^{\rm A}$ compared with the NMC \cite{Arneodo:1995cs,Amaudruz:1995tq} and E772 \cite{Alde:1990im} data.}
\label{Fig:RF2A1}
\end{figure}

As the DIS and DY data constitute our main data constraints, we display in Fig.~\ref{Fig:RF2A1} some examples of the measured nuclear modifications with respect to Deuterium,
$$
R_{F_2}^{\rm A}(x,Q^2) \equiv  \frac{F_2^A(x,Q^2)}{F_2^d(x,Q^2)},\qquad 
 R_{\rm DY}^{\rm A}(x_{2},M^2) \equiv \frac{\frac{1}{A}d\sigma^{\rm pA}_{\rm DY}/dM^2dx_{2}}{\frac{1}{2}d\sigma^{\rm pd}_{\rm DY}/dM^2dx_{2}}_{\Big|x_{2} \equiv \sqrt{M^2/s}\,e^{-y}}
$$
for different nuclei and compare with the EPS09.
The shaded blue bands always denote the uncertainty derived from the EPS09 error sets. Their size is comparable to the experimental errors, supporting our choice for $\Delta \chi^2$. 
\begin{figure}[!htb]
\centering
\vspace{-0.5cm}
\includegraphics[scale=0.20]{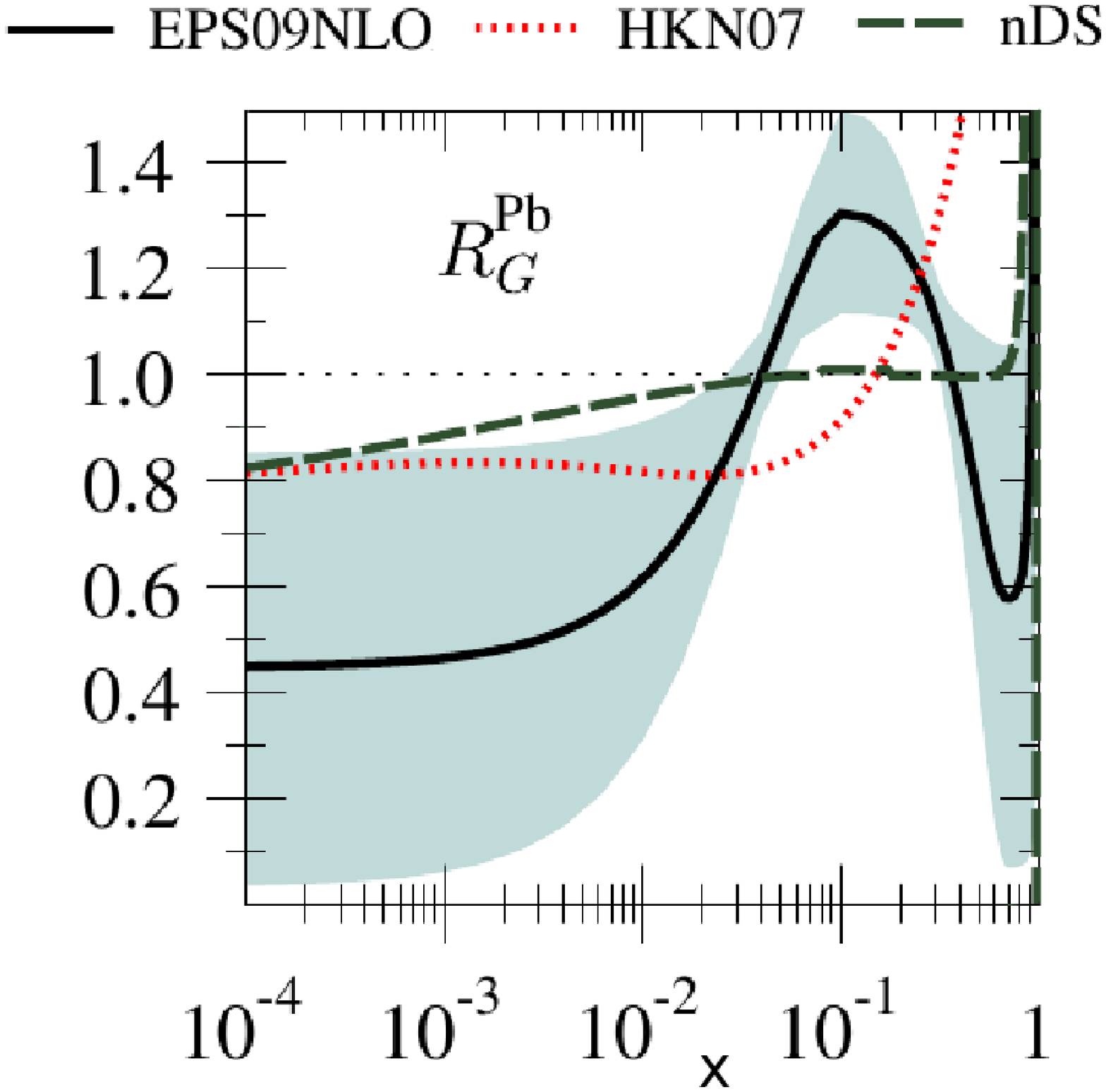}
\vspace{-0.3cm}
\includegraphics[scale=0.34]{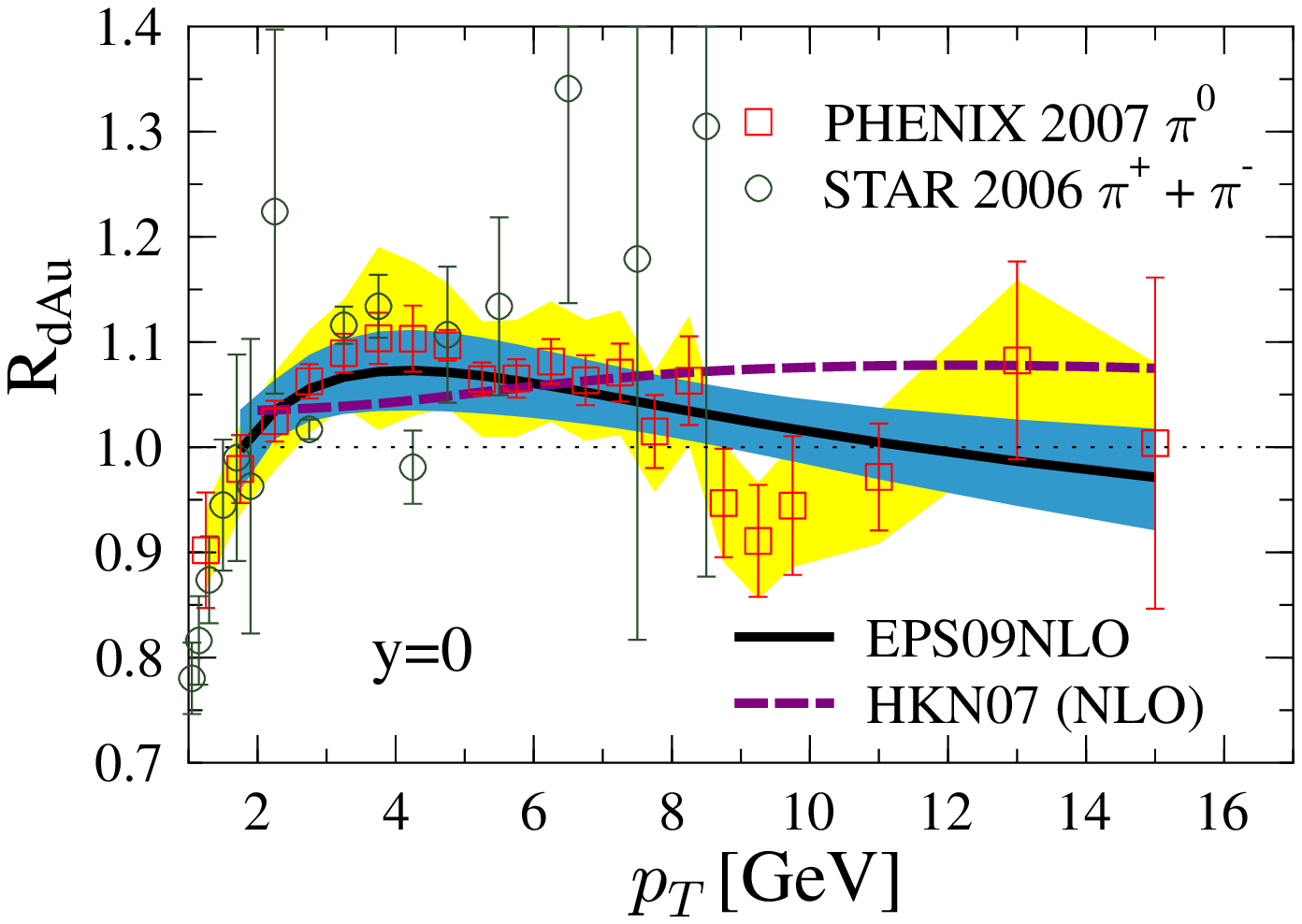}
\caption{Left: The modifications $R_G^{\rm Pb}$ at $Q^2=1.69 \, {\rm GeV}^2$ from
HKN07~\cite{Hirai:2007sx}, nDS~\cite{deFlorian:2003qf} and this work, EPS09 \cite{Eskola:2009uj}.
Right: The computed $R_{\rm dAu}$ for $\pi^0$ yield compared with the PHENIX \cite{Adler:2006wg} and 
STAR \cite{Adams:2006nd} data multiplied by $f_N = 1.03$ and $f_N = 0.90$ respectively.}
\label{Fig:PHENIX}
\end{figure}
\begin{figure}[!htb]
\center
\vspace{-0.3cm}
\includegraphics[scale=0.20]{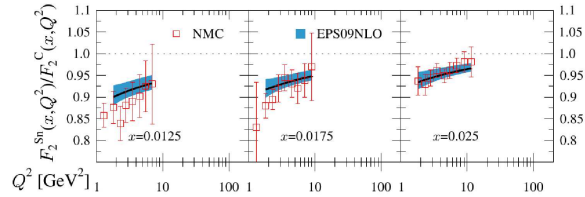}
\caption{The calculated scale evolution of the ratio $F_2^{\mathrm{Sn}}/F_2^{\mathrm{C}}$ compared with the NMC data \cite{Arneodo:1996ru}.}
\label{Fig:RF2_slopes}
\end{figure}
The nuclear modification for pion production is \nolinebreak defined \nolinebreak  as
$$
R_{\rm dAu}^{\pi}  \equiv  \frac{1}{\langle N_{\rm coll}\rangle} \frac{d^2 N_{\pi}^{\rm dAu}/dp_T dy}{d^2 N_{\pi}^{\rm pp}/dp_T dy} \stackrel{\rm min. bias}{=} \frac{\frac{1}{2A} d^2\sigma_{\pi}^{\rm dAu}/dp_T dy}{d^2\sigma_{\pi}^{\rm pp}/dp_T dy},
$$
where $\langle N_{\rm coll}\rangle$ denotes the number of binary nucleon-nucleon collisions and $p_T,y$ the pion's transverse momentum and rapidity. The comparison with the PHENIX and STAR data plotted in Fig.~\ref{Fig:PHENIX}, shows that the shape of the spectrum --- which in our calculation is a reflection of the similar shape in $R_G$ --- gets well reproduced by EPS09\footnote{The shape is practically independent of the set of contemporary fragmentation functions used.}. Figure~\ref{Fig:PHENIX} also presents a comparison of the EPS09 gluon modifications $R_G^{\rm Pb}$ with the earlier NLO analyses. The significant scatter of the curves highlights the difficulty of extracting the nuclear modifications from the DIS and DY data alone. Consequently, also the predictions for pion $R_{\rm dAu}$ differ significantly as is easily seen in Fig.~\ref{Fig:PHENIX}. This is actually good news as this type of data, especially with better statistics, may eventually discriminate between different proposed gluon modifications. 

Attention should be also paid to the experimentally observed scaling-violations and that the DGLAP
dynamics is able to reproduce them well. Such effects are most transparent e.g. in the small-$x$ structure
function ratios versus $Q^2$, of which Fig.~\ref{Fig:RF2_slopes} shows an example. To summarize, the excellent
agreement with the experimental data, $\chi^2/N \approx 0.79$, and especially the correct description
of the scale-breaking effects --- we argue --- is evidence for the applicability of collinear
factorization in nuclear environment. The best fit and all the 30 NLO nPDF error-sets are
available for practical use from \cite{EPS09code}. Also the leading-order EPS09 sets are provided.

\vspace{-0.5cm}
\section*{Acknowledgments}
\vspace{-0.2cm}
We thank the V., Y., \& K. V\"ais\"al\"a foundation, the M. Ehrnrooth foundation, and the Academy of Finland for
financial support.

 % do not change 
\end{document}